\documentclass[preprint,showpacs,preprintnumbers,amsmath,amssymb]{revtex4}
\usepackage{graphicx}
\usepackage{dcolumn}
\usepackage{bm}

\begin{document}
\title{Path Integral Ground State  with a Fourth-Order  Propagator: Application to Condensed Helium}
\author{Javier E. Cuervo, Pierre-Nicholas Roy,\footnote{pn.roy@ualberta.ca}}

\affiliation{Department of Chemistry, University of Alberta, Edmonton, Alberta, Canada, T6G 2G2}
\author{Massimo Boninsegni}
\affiliation{Department of Physics, University of Alberta, Edmonton, Alberta, Canada, T6G 2J1}
\date\today

\begin{abstract}
Ground state properties of condensed Helium are calculated using the Path Integral Ground State (PIGS)  method. A fourth-order approximation 
is used as short (imaginary) time propagator. We compare our results with those obtained with other Quantum Monte Carlo  techniques and different propagators. For this particular application, we find that the fourth-order propagator  performs comparably to the pair product approximation, and is far superior to the primitive approximation.  Results obtained for the equation of state of condensed Helium show that PIGS compares favorably to other QMC methods traditionally utilized for this type of calculation.
\end{abstract}

\maketitle

\section{INTRODUCTION}
\label{intro}
The Path Integral Ground State (PIGS) method\cite{Ceperley1,Magro} (alternatively known as Variational Path Integral) has emerged as an interesting alternative to other, numerically exact  ground state Quantum Monte Carlo (QMC) methods  such as Green function (GFMC) and Diffusion Monte Carlo (DMC), which have been widely used over the past two decades.\cite{notex}

The basic ideas of PIGS are common to the other projection techniques. Consider for definiteness a system of $N$ identical particles of mass $m$;  the quantum-mechanical Hamiltonian $\hat {\cal H}$ 
of the system is 
\begin{equation}\label{ham}
\hat {\cal H}=\hat {\cal H}_\circ + \hat V = -\lambda \sum_{i=1}^N\nabla^2_i + V(R) \end{equation}
where $\lambda={\hbar^2}/{2m}$, $R\equiv {\bf r}_1{\bf r}_2...{\bf r}_N$, are the positions of the $N$ particles, and $V(R)$ is the total potential energy of the system associated with the many-particle configuration $R$ (this is typically the sum of pairwise interactions, but can be more general). The exact ground state wave function $\Phi_\circ(R)$   can be formally  obtained from an initial trial wave function $\Psi_T(R)$ as
\begin{eqnarray}\label{proj}
\Phi_\circ(R) \propto {\rm lim}_{\beta\to\infty} \int dR^\prime\ G(R,R^\prime,\beta) \ \Psi_T(R^\prime)
\end{eqnarray}
where \begin{equation}\label{prop}
G(R,R^\prime,\beta) = \langle R | {\rm exp}[-\beta\hat{\cal H}]|R^\prime\rangle
\end{equation}
is commonly referred to as the imaginary-time propagator. While Eq. (\ref{proj}) is formally exact, for a nontrivial many-body problem one does not normally know how to compute $G(R,R^\prime,\beta) $ exactly. However,  using one of several available schemes, it is possible to obtain approximations for $G$, whose accuracy increases as $\beta\to 0$;  if $G_\circ(R,R^\prime,\beta)$ is one such approximation, one can take advantage of the identity ${\rm exp}[-\beta\hat H]\equiv \biggl ({\rm exp}[-\tau\hat H]\biggr )^M$, with $\beta=M\tau$, and obtain  $G(R,R^\prime,\beta)$ as 
\begin{eqnarray}\label{path}
G(R,R^\prime,\beta) \approx \int dR_1...dR_{M-1} G_\circ(R,R_1,\tau) \ G_\circ(R_1,R_2,\tau)...G_\circ(R_{M-1},R^\prime,\tau)
\end{eqnarray}
For finite $M$, $\tau$, Eq. (\ref{path}) is approximate, becoming exact only  in the limit $M\to\infty$ (i.e., $\tau\to 0$). In a numerical calculation, one must necessarily work with finite values of $M$ and $\tau$; $\tau$ must be chosen sufficiently small, so that the replacement of $G$ by $G_\circ$ entails no significant loss of accuracy, whereas the product $M\tau$ should be large enough so that the formally exact $\beta\to\infty$ limit can be approached to the desired precision. 

Thus, regardless of the numerical scheme adopted to implement Eq. (\ref{path}),  it is clearly advantageous to work with a  ``short-time" approximation $G_\circ$ which will allow one to obtain reliable ground state estimates without having to resort to impractically large values of $M$. 

In the simplest approximation, known as {\it primitive}, one replaces $G$ by $G_P$,  given by
\begin{equation}\label{eq1}
G_{P}(R,R^\prime,\tau)=\rho_F(R,R^\prime,\tau)\  e^{-\frac{\tau}{2}[V(R)+V(R^\prime)]} ),
\end{equation} where
\begin{eqnarray}
\rho_{F}(R,R^\prime,\tau)\equiv
\langle R|{\rm exp}[-\tau\hat{\cal H}_\circ]|R^\prime\rangle= (4\pi\lambda\tau )^{-3N/2}
\ \prod_{i=1}^N {\rm exp}\biggl [-\frac{({\bf r}_{i}-{\bf r}^\prime_{i})^2}{4\lambda\tau}\biggr ]
\end{eqnarray}
is the exact propagator for a system of non-interacting particles. The primitive approximation is accurate up to terms of order $\tau^3$.

The primitive approximation (PA), when used in Eq. (\ref{path}),  leads to an expression for the propagator $G$, which is accurate to within a term of order $1/M^2$. It has been traditionally the most common choice for ground state calculations based on DMC, normally with the incorporation of techniques such as importance sampling and rejection,\cite{reynolds,umrigar}  which have been demonstrated to improve its performance significantly. 

A number of  propagators, enjoying higher order of accuracy in $\tau$ have been derived\cite{chin2001,Takahashi,Li} but their use in DMC calculations has been relatively rare.\cite{bressanini,imbecile}

However, this issue has gained renewed attention in recent times, as PIGS has elicited significant interest as a ground state method potentially superior to DMC. 
PIGS  has the advantage of providing relatively easily expectation values for physical observables that do not commute with the Hamiltonian operator. Moreover, PIGS is immune from the bias affecting DMC, arising from the fact that one is working with a  finite population of random walkers. Although a trial wave function $\Psi_T$ for the physical system of interest is required in PIGS calculations (as in DMC), some empirical evidence suggests \cite{Magro} that results for observables other than the energy are considerably less sensitive to the choice of $\Psi_T$ than in DMC. Last but not least, PIGS can be relatively easily generalized  to compute expectation values of off-diagonal operators, such as the one-body density matrix,\cite{moroni04} whose computation can only be carried out approximately within DMC.

Because PIGS is much closer, both in spirit and numerical implementation,  to finite temperature Path Integral Monte Carlo (PIMC) than to DMC, an obvious choice for the short imaginary time propagator to use in PIGS calculations is the pair product approximation (PPA) of Pollock and Ceperley,\cite{Ceperley1} which has proven to boost enormously the efficiency of PIMC calculations for hard-sphere-like quantum many-body systems, such as condensed Helium. On the other hand, the PPA can prove quite cumbersome to use, and its effectiveness for systems other than Helium has not yet been quantitatively established. It is  therefore desirable to have an alternative to the PPA, easier to implement in practice and affording a substantial improvement over the primitive approximation.

The goal of this work is to assess the performance of PIGS, in combination with a recently proposed fourth-order propagator (FOP).\cite{moron,voth} Such a propagator has been shown\cite{voth,boninsegni04a} to improve significantly the convergence of PIMC calculations, with respect to the primitive approximation; it has also been used in recent studies, based on PIGS, of adsorbed Helium films.\cite{boninsegni04b} Our aim is to carry out a systematic comparison of different PIGS implementations based on different propagators, as well as of PIGS and DMC. 

The many-body system of choice to carry out our study  is  condensed $^4$He, which is the {\it de facto} accepted testbench for ground state Quantum Monte Carlo calculations, at least for Bose systems. Because it is a strongly correlated system, the study of the ground state properties of condensed Helium is clearly a cogent test of  {\it any}  many-body computational tool; it is particularly suited  for a comparative study of QMC methods, because of the large body of work on Helium, based on different QMC techniques,  that has been carried out over the past thirty years. 

The remainder of this paper is structured as follows: in the next section we describe the model physical system for which calculations have been carried out; 
in Sec. \ref{theo} we briefly review the theory of PIGS, and illustrate in detail the propagator used in this work. In Sec. \ref{results} we first present results for the ground state of liquid $^4$He at the equilibrium density, and assess the convergence of the numerical estimates for various physical quantities, both with respect to the imaginary time step $\tau$ and to the total projection time $\beta=M\tau$.  We then present results for the zero temperature equation of state of condensed $^4$He, computed by PIGS,  in an extended density range, and compare our results to those obtained by DMC and GFMC. We outline our conclusions in  Sec. \ref{conclusions}.    

\section{Model}\label{mod}

We model condensed $^4$He as an ensemble of $N$ $^4$He atoms, regarded as point particles,
moving in three dimensions. The system is enclosed in a vessel shaped as a parallelepiped of volume $\Omega$, with periodic boundary conditions in all directions. The nominal density of the system is $\rho=N/\Omega$.
The quantum-mechanical many-body Hamiltonian is given by Eq. (\ref{ham}), with the following choice for the potential energy $V(R)$:
\begin{equation}\label{one}
V(R) = \sum_{i<j} v(r_{ij}) 
\end{equation}
Here, $v$ is the potential describing the interaction between two helium atoms,
only depending on their relative distance. Although three-body terms are known to be important, if one is to achieve an accurate description of the equation of state of condensed Helium at low temperature, \cite{pederiva} we limit ourselves to pair potentials only, since that is how most of the previous calculations have been carried out, and our primary aim here is compare with existing calculations.

The determination of a potential to describe the interactions between a pair of Helium atoms has been the goal of a long-lasting research effort. 
Aziz and collaborators{\cite {aziz79}} carefully combined theoretical 
and experimental gas phase data on the interaction of two atoms, 
to obtain a highly reliable pair potential, which has proven adequate \cite{Ceperley1}
to describe energetic and structural properties of condensed $^4$He. Over the course of the past two decades, several refinements of the original Aziz potential have been proposed. However,  the early version\cite{aziz79} (henceforth referred to as Aziz-I) has been used in most previous QMC studies, which is why most of the results presented in this paper are based on it. For the sake of comparison with more recent calculations, we also obtained in this work estimates with a different potential, heretofore referred to as Aziz-II.\cite{Aziz-95}

\section{Methodology}
\label{theo}

The Path Integral Ground State  method allows one to obtain numerical estimates, in principle exact,
of ground state expectation values for quantum many-body systems described
by a Hamiltonian such as (\ref{ham}). 
The prescription, which aims at implementing Eqs. (\ref{proj},\ref{prop}) numerically,  is based on the approximate equality (\ref{path}) and on probabilistic considerations. Specifically, this is how one implements it: One generates sequentially, on a computer, a large set $\{X^p\}$, $p=1,2,...,P$, of many-particle paths $X\equiv R_0 R_1\ ...\
R_{2M}$ through configuration space. Each $R_j \equiv {\bf r}_{j1}{\bf
r}_{j2} \ ... \ {\bf r}_{jN}$ is a point in 3$N$-dimensional space,
representing positions of the $N$ particles (i.e., $^4$He atoms) in the
system. These paths are statistically sampled from a probability density
\begin{eqnarray} 
{\cal P}(X)\propto \Psi_T(R_0)\Psi_T(R_{2M}) \
 \biggl \{ \prod_{j=0}^{2M-1} G_\circ(R_j,R_{j+1},\tau)\biggr \}\label{pippo}
\end{eqnarray}
where $\Psi_T(R)$ is a variational wave function for the ground state of
the system and $G_\circ(R,R^\prime,\tau)$ is the short-time approximation for the imaginary-time propagator $G$.

It is a simple matter to show \cite{Ceperley1,Magro} that in the limits $\tau
\to 0$, $M\tau \to \infty$, $R_{M}$ is sampled from a probability density
proportional to the square of the exact ground state wave function $\Phi_\circ(R)$, {\it irrespective of the choice of} $\Psi_T$.\cite{notey} One can therefore
use the set $\{R_M^p\}$ of ``midpoint" configurations $R_{M}$ of the statistically sampled paths, to compute ground state expectation values of thermodynamic quantities $F(R)$ that are diagonal in the position
representation, simply as statistical averages, i.e.
\begin{equation}\label{pix}
\langle\Phi | \hat F(R)|\Phi\rangle \approx \frac{1}{P}\ \sum_{p=1}^P\
F(R_M^p),
\end{equation} 
an approximate equality, asymptoptically exact in the $P \to
\infty$ limit. The ground state expectation value of the energy can be
obtained in several ways; it is particularly convenient to use the ``mixed estimate"
\begin{equation}\label{mix}
\langle\Phi_\circ|\hat H|\Phi_\circ\rangle \approx \sum_{p=1}^P\ \frac {\hat H\Psi_T(R_1^p)}{\Psi_T(R_1^p)}
\end{equation}
which provides an unbiased result for the Hamiltonian operator $\hat H$,
as this commutes with the imaginary time evolution operator exp$[-\tau\hat H]$ (note that $R_{2M}$
may just as well be used in (\ref{mix})). An alternate estimator can be obtained from the identity
\begin{equation}
\langle \Phi_\circ|\hat{\cal H}|\Phi_\circ\rangle = - {\rm lim}_{\beta\to\infty}\ \frac{\partial}{\partial\beta}\ 
{\rm log}\biggl \{ \int dR\ dR^\prime \Psi_T(R)\Psi_T(R^\prime)G(R,R^\prime,\beta)\biggr \}
\end{equation}
which results in the same energy estimator commonly used in PIMC calculations.\cite{Ceperley1}  While the latter estimator can be more robust, as it is less dependent on the choice of trial wave function, the mixed estimator has typically a much lower variance.

Because $M$ is necessarily finite, for a given value of $\tau$ one must
repeat the calculation for increasing $M$, until convergence of the
estimates is achieved, within the desired accuracy. For any finite $M$,
the energy expectation value is a strict upper bound for the exact value
(hence the alternate name {\it Variational} Path Integral).
Numerical extrapolation
of the estimates obtained for different $\tau$ must then be carried out,
in order to obtain results in the $\tau\to 0$ limit.

Next, we discuss our choice of short imaginary time propagator $G_\circ$. As mentioned in the Introduction, several forms are possible for $G_\circ$; it must be clarified at the outset that the issue here  is merely one of {\it computational efficiency}. For, a more accurate form for $G_\circ(R,R^\prime,\tau)$ (or, a more accurate trial wave function $\Psi_T$) will allow one to observe convergence with a smaller value of $M$ and/or a greater time step $\tau$, but will not otherwise affect the results, provided enough computer time.
In this work, the following approximation was used:
\begin{equation}\label{voth}
G_\circ(R_j,R_{j+1},\tau)= \rho_F(R_j,R_{j+1},\tau) 
\ {\rm exp} \biggl [-\frac{2\tau V(R_{j})} {3}\biggr ]\
\rho_V(R_j)
\end{equation}
where 
 
\begin{eqnarray} \rho_V(R_{j}) = 
{\rm exp}\biggl [-\frac{2\tau{V}(R_{j})}{3}-\frac{\tau^3\hbar^2} {9m}\sum_{i=1}^N
(\nabla_{i}{V}(R_{j}))^2\biggr ]
\end{eqnarray}
if $j$ is odd, whereas $\rho_V(R_j)=1$ is $j$ is even. This is a particular case of a more general propagator, for which it can be shown\cite{moron} that  $G(R,R^\prime,\tau)=G_\circ(R,R^\prime,\tau)+{\cal O}(\tau^5)$.

For a given choice of $M$ and $\tau$, one computes the approximate estimates (\ref{pix},\ref{mix}) by generating the set $\{X^p\}$ by means of a random
walk through path space, using the Metropolis algorithm. The same path
sampling techniques utilized in finite temperature PIMC can be used in VPI.
In this work, multilevel sampling with bisection and staging\cite{Ceperley1, Magro} was adopted, together with rigid displacements of entire single-particle
paths. In all of these moves, the proposed new positions are sampled based on the free-particle part $\rho_F$  of the short-time propagator (\ref{voth}), as the rest of the propagator enters in the acceptance/rejection step only.
The only difference between multilevel
moves for central portions of the path and for those including the ends
(i.e., ``slices" 1 and 2$M$) is the presence, in the latter, of the trial
wave function $\Psi_T$ in the Metropolis acceptance/rejection test. Other
strategies have been proposed, allowing one to update paths in the vicinity
of the ends, e.g., ``reptation" type moves;\cite{baroni98} however, in this
work we have not made use of them.

The main difference between DMC and PIGS is the fact that DMC implements the imaginary time evolution of the initial, trial state $\Psi_T$, by means of a guided random walk through configuration space of a population of independent random walkers. An essential ingredient of this approach is the fact that walkers, along the random walk, accumulate weights proportional to ${\rm exp}[-\int d\tau E_L(\Psi_T(R),\tau)$, where 
$E_L(\Psi_T(R))\equiv\hat{\cal H}\Psi_T(R)/\Psi_T(R)$ is the local energy given by the trial wave function 
at the configuration $R$, visited at imaginary time $\tau$ by a given walker. Typically, weights fluctuate considerably, both along the random walks, as well as within the population at any given time. Therefore, it proves convenient to reconfigure the population, every now and then during the calculation, so that walkers whose weights have become negligibly small are discarded, and copies are made of walkers whose weights are larger. This reconfiguration, known as {\it branching}, is done in such a way that the size of the population remains constant; it has been shown to improve considerably the efficiency of the algorithm (see, for instance, Ref. \onlinecite{umrigar} for details).

The main advantage of this computational strategy, at least in principle, is that the projection time can be made very large with little computational effort. On the other hand,  a bias is introduced in the computation, as one must necessarily work with a finite population of walkers (in order for the algorithm to be exact, such a population should be infinite). There has been little work aimed at establishing the magnitude of such a bias on the computed expectation values, but some calculations have shown that  it can be significant, particularly when trying to estimate expectation values of operators that do not commute with the Hamiltonian.\cite{runge92,calandra}

Within PIGS, on the other hand, there is no such bias, as there is no population and no branching is needed. The possible drawback is that one is working with a finite projection time. While in principle such a projection time can be made large, there is a potential problem with rendering the paths too long (i.e., working with too large a number $M$), as long paths diffuse through configuration space increasingly slowly, and therefore one might face an ergodicity problem. However, very little experimentation has been carried out so far, and therefore the real importance of this issue has not yet been assessed quantitatively.

We conclude this section by noting another important difference between DMC and PIGS (at least the way it has been implemented so far, including in this work): there is no importance sampling in PIGS, i.e., no information about the trial wave function is included in the sampling of paths through configuration space (with the exception of the end slices). Although one may expect this to result in significantly greater statistical uncertainties when using PIGS, the results presented here do not seem to support this speculation.
 
\section{RESULTS}
\label{results}
The PIGS calculations whose results we describe below were performed on systems with a number of particles ranging from 256 to 288. All simulations are started with particles arranged on a regular lattice, with the same set of positions for all slices.
The trial wave function utilized is of the Jastrow type, namely 
\begin{equation}\label{jastrow}
\Psi_T(R)={\rm exp}\biggl [-\frac {1}{2}\sum_{i<j} u(|{\bf r}_i-{\bf r}_j|)\biggr ]
\end{equation}
Most of the calculations were performed with the following form for the pseudo-potential $u$
\begin{equation}\label{pseudo}
u(r)=\frac{\alpha}{1+\beta r^5}
\end{equation}
with $\alpha=19$ and $\beta=0.12$ \AA$^{-5}$. These values of the parameters were found by minimizing the 
expectation value of the energy obtained with $\Psi_T$, in a separate variational calculation. For comparison, we also performed some of the calculations using the McMillan pseudo-potential $u(r)=b/r^5$, with $b$=3.07 \AA. As expected, the results obtained with the two different trial wave functions are in agreement, with statistical uncertainties.

Our estimates for the potential energy include a contribution due to particles outside the simulation cell; we estimated such a contribution by assuming that the pair correlation function $g(r)=1$ outside the cell. Because of the relatively large number of particles in our simulated systems, this should be an excellent approximation.

All of our energy estimates were obtained using the mixed estimator (Eq. (\ref {mix})); the potential energy per particle, on the the hand, as well as the pair correlation function $g(r)$, is computed as indicated in Eq. (\ref{pix}), whereas the kinetic energy is obtained by subtracting the potential from the total energy.

\subsection{Time Step}\label{tst}
\begin{figure}[ht]
\includegraphics[width=\textwidth]{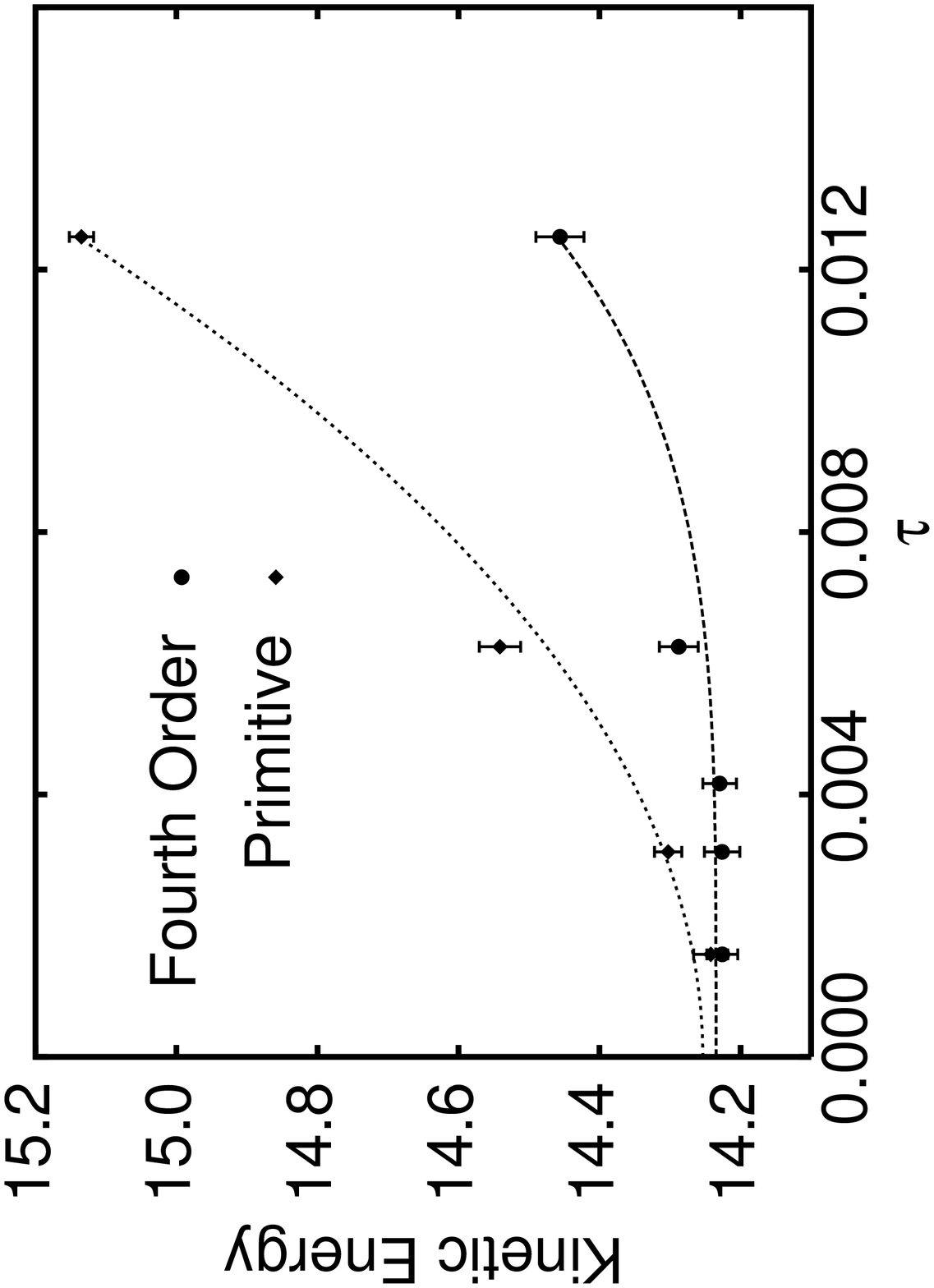}
\caption{}
\label{fig1}
\end{figure}

We discuss first the dependence of different expectation values calculated by PIGS on the imaginary time step $\tau$. We begin with the ground state energetics. Figure \ref{fig1} shows estimates (circles) of the kinetic energy $K$ per atom of liquid $^4$He at the equilibrium density $\rho_e=0.02186$ \AA$^{-3}$, for  different choices of the time step $\tau$ (in K$^{-1}$). The potential used in this calculation is the Aziz-I.  These results are obtained for a total projection time $\beta=0.25$ K$^{-1}$, and with the FOP of Eq. (\ref {voth}).
Figure \ref{fig1} also shows the same quantity computed in a separate PIGS calculation, based on the PA (diamonds). Calculations based on the two different approximations converge, as they ought to, to the same value in the $\tau\to 0$ limit, within statistical uncertainties. However, the low-$\tau$ behavior is different in the two cases.

It can be shown\cite{Brualla} that in the $\tau\to 0$ limit expectation values computed by PIGS behave as follows:
\begin{equation}\label{eq3}
K(\tau)=K_\circ+A_\delta(\tau)^\delta+A_{\delta+2}(\tau)^{\delta+2}
\end{equation} 
where $K_\circ$ is the  exact ground state result, and $\delta$ depends on the accuracy of the approximation used for $G_\circ$;  for the PA, $\delta$=2, whereas if the FOP is used one has $\delta$=4. 

In principle, in order to estimate $K_\circ$ one ought to obtain estimates of $K(\tau)$ for a number of significantly different values of $\tau$,  and extrapolate the value of $K_\circ$ by fitting the data using (\ref{eq3}). However, while this cumbersome procedure is inevitable if one uses the PA, given the slow convergence of the estimates, Fig. \ref{fig1} clearly shows that values of $K(\tau)$ obtained with the FOP are indistinguishable, within statistical uncertainties, for $\tau\le\tau_\circ\approx 0.005$ K$^{-1}$. The same analysis can be carried out for other quantities, such as the potential and the total energy per particle; in all of these cases, we found the expected quartic behavior of the estimates as a function of $\tau$, with the same value of $\tau_\circ$, namely 0.005 K$^{-1}$, below which estimates no longer change, within statistical uncertainties.

This value of time step is a little over a factor two smaller than that (0.0125 K$^{-1}$) used in PIGS calculations for $^4$He based on the PPA.\cite{Magro}  In other words, even though the PPA allows one to observe convergence of the estimates with a greater time step (and therefore fewer slices), the improvement afforded by the use of the PPA over the FOP, in the context of PIGS calculations for condensed $^4$He, is not nearly as large as that observed  in finite temperature calculations, where the PPA can reduce the number of time slices needed to achieve convergence at low temperature ($T \le $ 2 K), with respect to the PA, by as much as a factor of 50 or greater (see Ref. \onlinecite{Ceperley1}).
In our view, given the computing facilities commonly available nowadays, the significant computational simplification arising from the use of the FOP more than compensates for the factor of two more imaginary time slices needed, with respect to calculations based on the PPA.\cite{notew}

Finally, we note that the optimal value $\tau_\circ$, found here for the time step, is as much as ten times greater than that required in DMC calculations based on the PA.\cite{boronat94,notezz}  It is also approximately five times greater than that used in Reptation Quantum Monte Carlo.\cite{baroni98} Although we are comparing here calculations using slightly different versions of the Aziz potential, the optimal value of the time step needed is largely insensitive to the fine details of the interaction.

\subsection{Projection Time $\beta$}\label{proje}

\begin{figure}[ht]
\includegraphics[width=\textwidth]{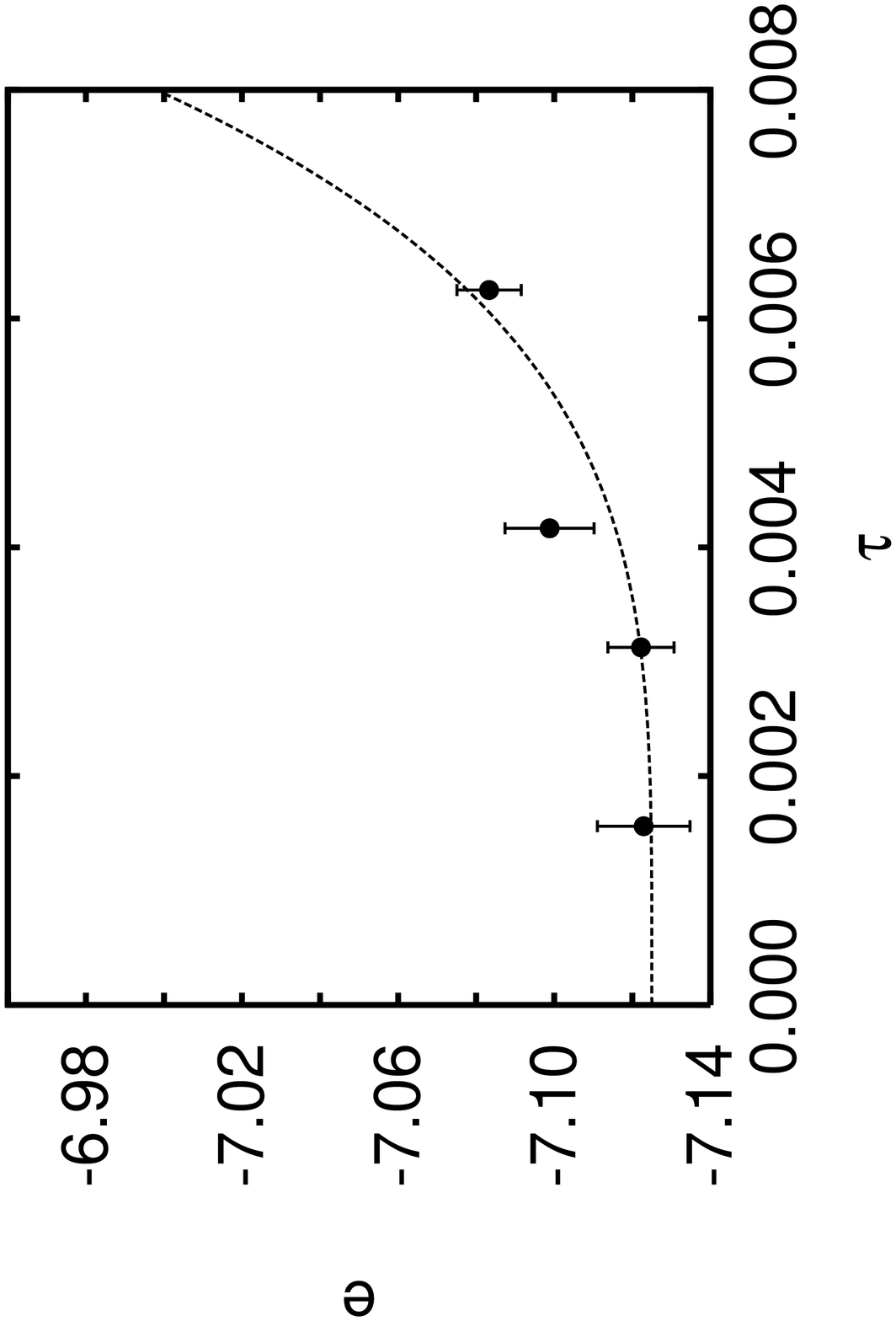}
\caption{}
\label{fig2}
\end{figure}

Figure \ref{fig2} shows our PIGS energy estimates for different values of $\tau$ at the equilibrium density, for a projection time $\beta$=0.25 K$^{-1}$. The extrapolated $e_\circ\equiv e(\tau=0)$ value is $e_\circ=-7.123\pm$0.003 K, which is in perfect agreement with the most recent DMC result\cite{boronat94} based on the same (Aziz-I) potential. 
This suggests that, although PIGS estimates for the total energy are strictly variational,  a projection time $\beta=0.25$ K$^{-1}$ is sufficiently long to obtain accurate ground state results, at least at this density and with the trial wave function utilized. 

\begin{figure}[ht]
\includegraphics[width=\textwidth]{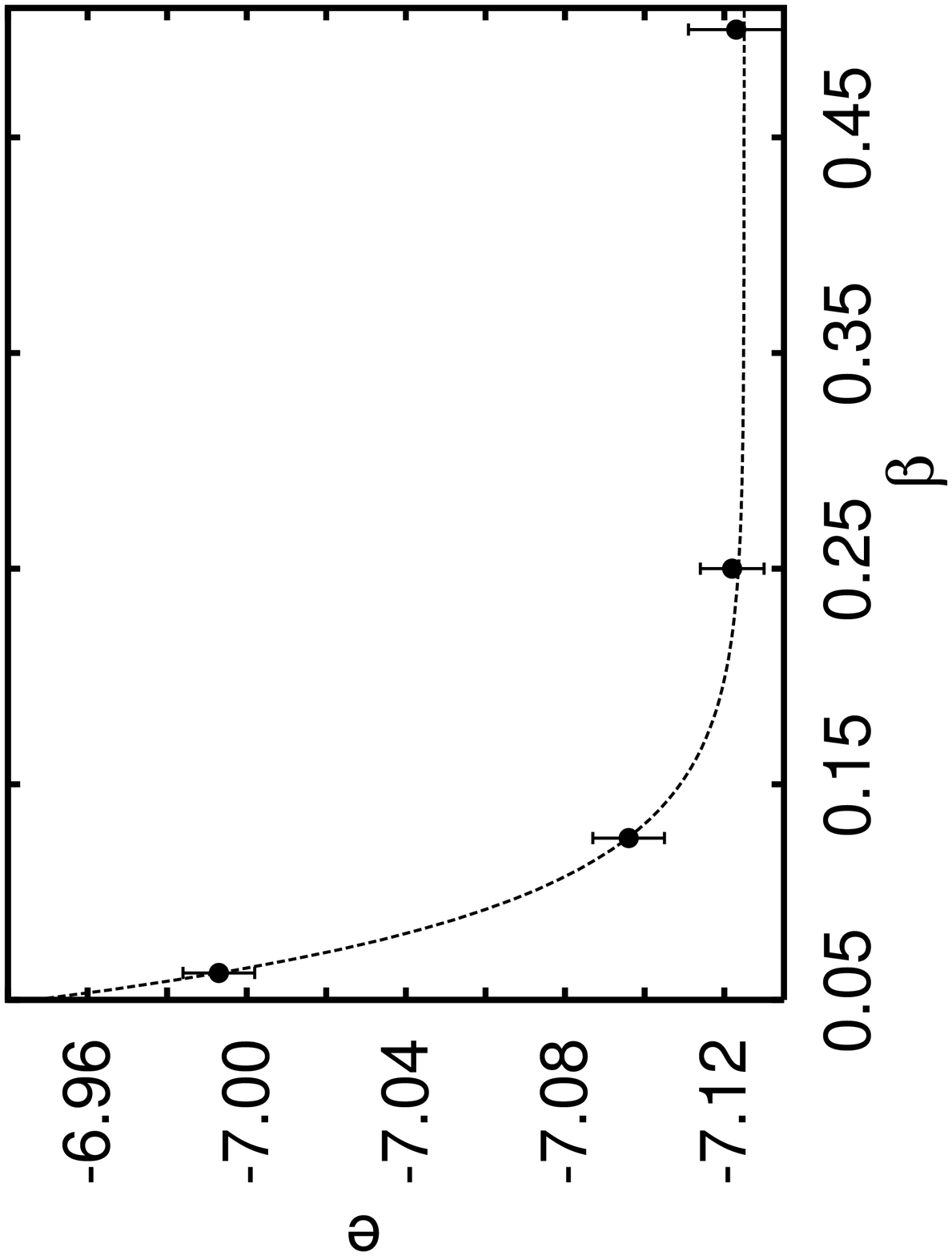}
\caption{}
\label{fig3}
\end{figure}

In order to establish this conclusion more quantitatively, we computed the energy expectation value $e(\beta,\tau)$ for different projection times $\beta$ (specifically,  0.0625 K$^{-1}$ $\le \beta\le $ 0.5 K$^{-1}$), with a fixed value of the time step $\tau=3.125\times10^{-3}$ K$^{-1}$. These results are shown in Fig. \ref{fig3}. 

It is simple to show\cite{caffarel92} that, in the limit $\beta\to\infty$, the true ground state expectation value must be approached exponentially, i.e., $e(\beta)\sim e_\circ + b\ {\rm exp}(-c\beta)$, where the $c$ is essentially the energy gap between the ground state and the first excited state. This simple expression provides an excellent fit to our energy estimates $e(\beta,\tau)$, as shown in Fig. \ref{fig3}. Similar results are obtained for the kinetic and potential energy, confirming that a projection time $\beta\approx0.25$ K$^{-1}$ is sufficiently long to extract accurate energetics, at least at the equilibrium density and with the trial wave function utilized.

The behavior of structural quantities (such as the pair correlation function) vis-a-vis the time step $\tau$ and the projection time $\beta$ is more difficult to assess quantitatively. In general, within our statistical uncertainties our results for $g(r)$ do not change for $\beta \ge 0.125$ K$^{-1}$, i.e., the projection time required to observe convergence of $g(r)$ is about half of that needed for the energy.  Moreover, although we did not pursue this aspect quantitatively, our observation  is that a greater value of the time step can be used, in order to obtain accurate estimates of $g(r)$, than that needed for the energy. In other words, once $\beta$ and $\tau$ are adjusted to yield satisfactory energy results, one can be reasonably confident that expectation values of structural quantities will  also be accurate. Results for $g(r)$ are shown in the next subsection.

\subsection{Comparison with other calculations}
In order to carry out a more thorough comparison of our results with those of analogous calculations, we computed energetics of liquid $^4$He at the equilibrium density using a different Aziz potential, namely the Aziz-II. This is the same potential utilized in Ref. \onlinecite{Magro}. Results for the kinetic and total energy per atom, obtained in this work, as well as in Ref. \onlinecite{Magro} (also using PIGS but with the PPA) and with DMC (Ref. \onlinecite{casulleras}) are shown in Table \ref{tab1}. Within the quoted statistical uncertainties, all calculations are in agreement as far as the energy per particle is concerned. As for the kinetic energy, it is only evaluated approximately within GFMC, based on an extrapolation procedure that retains some of the bias associated to the initial trial wave function utilized; thus, it is not surprising that the GFMC estimate disagrees with the PIGS ones. The comparison of the two PIGS  estimates of the kinetic energy seems satisfactory, given the relatively large statistical error quoted in Ref. \onlinecite{Magro} (see, however, our comment in Ref. \onlinecite{notew}). 

\subsection{Equation of State of Condensed $^4$He}
In order to provide a fuller assessment of PIGS,  and carry out an extended comparison with other methods, we computed the equation of state of condensed $^4$He in the density range 0.0196 \AA$^{-3}$ $\le\rho\le$ 0.0292 \AA$^{-3}$, i.e., up to the melting density. We used the Aziz-I potential, and set the projection time $\beta=0.25$ K$^{-1}$ and the time step $\tau=1.5625\time 10^{-3}$ K$^{-1}$, which corresponds to $M=160$ (based on the results shown in Sec. \ref{tst}, one could argue that a larger time step could have been used as well; however, because CPU time was not an issue for these calculations, we opted for a safer choice, so as to avoid the need for any $\tau$ extrapolation). 

 Our calculations were carried out for systems with 256 particles for $\rho \le 0.026$ \AA$^{-3}$, whereas for greater densities we used 288 particles, arranged on a regular hcp lattice at the beginning of the simulation. Each thermodynamic point  requires a few days of CPU time on a common workstation.

The significance of this calculation lies in the fact that we used the same Jastrow wave function (\ref{jastrow}), with the pseudo-potential $u$ given by (\ref{pseudo}) and with the same optimal parameters obtained at the equilibrium density, {\it for all values of the density considered}. In other words, even though the density range studied includes regions where the equilibrium phase is a coexistence of liquid and solid ($\rho > 0.0262$ \AA$^{-3}$), or a solid only (at the highest density), we always utilized a trial wave function that does not break translational invariance.\cite{notez} The purpose of such a choice is precisely that of assessing the robustness of the algorithm, namely its predictive capability when the physics of the ground state is qualitatively different than that described by the trial wave function. 

The projection time utilized, as discussed in \ref{proje}, is sufficiently long to ensure convergence to the ground state at the equilibrium density; at higher densities, one may expect a longer projection time to be required, as the trial wave function (\ref{jastrow}) provides less and less accurate a description of the ground state physics. At any rate, $e(\beta,\tau\to 0)$ is always a {\it strict upper bound} on the exact ground state energy for finite $\beta$. Thus, our energy estimates can always regarded as variational results, upon which one may improve arbitrarily by increasing $\beta$.

\begin{figure}[ht]
\includegraphics[width=\textwidth]{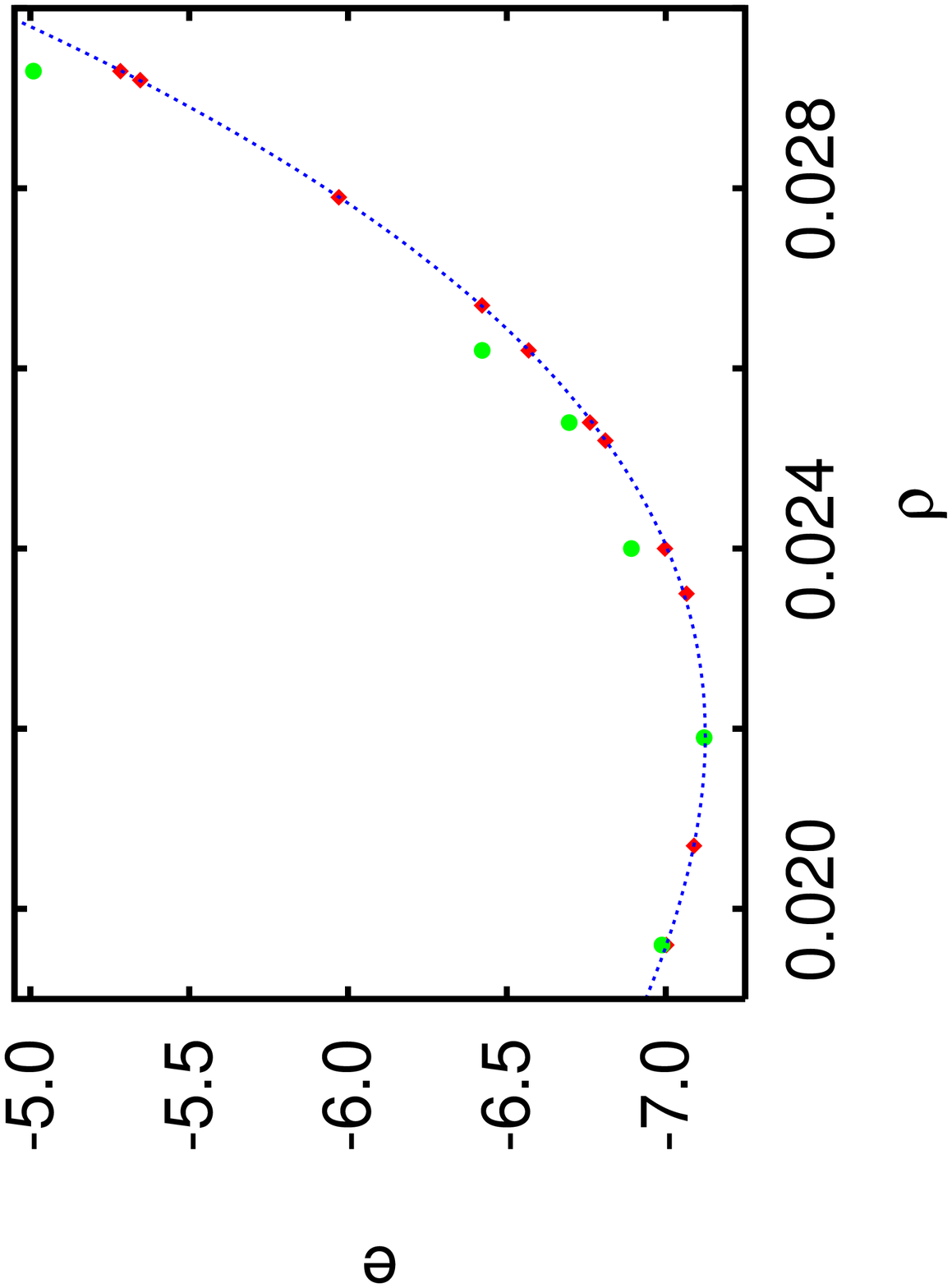}
\caption{}
\label{eos}
\end{figure}

Our estimates for the total energy per particle $e(\rho)$ are given in Table \ref{tab2} and shown in Fig. \ref{eos}. We compare them to GFMC (Ref. \onlinecite{kalos}) and DMC results (Ref. \onlinecite{boronat94}). Besides being the most recent (at least to our knowledge) for the potential used here, these DMC results are particularly appropriate for this comparison, as they were also obtained using a translationally invariant Jastrow  trial wave function. 

While there is quantitative agreement among all calculations at the equilibrium density and below, there are significant deviations at higher densities. In particular, PIGS energy estimates fall consistently {\it below} the DMC ones. On the other hand, they appear to be  in reasonable agreement with the earlier GFMC results, although the relatively large statistical uncertainties affecting those GFMC calculations makes it difficult to establish this conclusion firmly. 

The difference between PIGS and DMC results for the energy increases with density. This seems noteworthy, as one might expect the limited projection time of a PIGS calculation to put it at a disadvantage with respect to DMC, particularly when the trial wave function does not capture the physics of the ground state (e.g., above the freezing density). Our results point instead to a greater robustness of PIGS, compared to DMC.
It is not immediately obvious what the reason for the discrepancy between PIGS and DMC energy estimates might be. Both calculations were performed for systems of relatively large size, i.e., finite-size corrections are expected to be much smaller than the differences in energy seen here. Nor does it seem likely that the slightly different Jastrow trial wave function utilized in the two calculations may be responsible. In our view, a possible explanation may indeed lie with the finite population size employed in the DMC calculations (a few hundred walkers). As mentioned above, the use of a finite population size, has the effect of introducing a bias in the estimates obtained within DMC. Such a bias must be corrected for, if one is attempting to obtain accurate energy results.\cite{calandra}

At the melting density $\rho=0.0293$ \AA$^{-3}$, our PIGS result is in agreement with the GFMC estimate of Ref. \onlinecite{kalos}, at least within the statistical uncertainties of the GFMC calculation, which are about ten times greater than ours. The GFMC result was obtained using a Jastrow-Nosanow trial wave function, which explicitly breaks translational invariance by incorporating a one-body term whose effect is that of ``pinning" atoms at specific lattice positions. Such a trial wave function has been shown to lead to more accurate estimates in variational calculations for the crystal phase; as mentioned above, however, our results based on PIGS, and on a translationally invariant, two-body Jastrow wave function, are consistent with GFMC estimates, showing that the variational bias arising from the use of a simple wave function is removed.

\begin{figure}[ht]
\includegraphics[width=\textwidth]{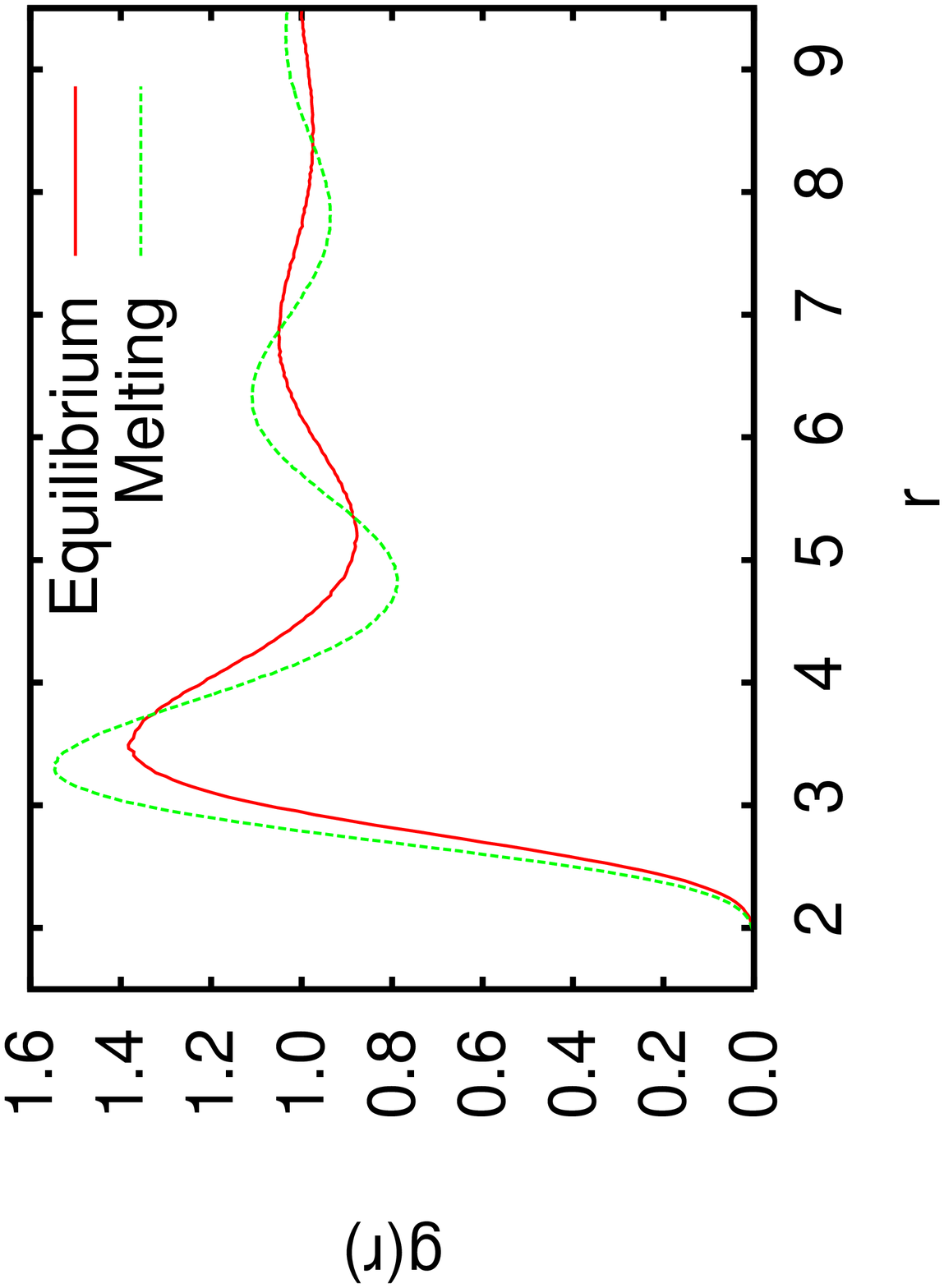}
\caption{}
\label{gofr}
\end{figure}

As previously mentioned, the starting configuration of all of our simulations is with particles arranged on a regular lattice, simple cubic for $\rho < 0.0262$ \AA$^{-3}$, hcp for higher densities. Obviously, in the course of the simulation particles do not remain at their initial lattice positions, nor do they necessarily continue to form a crystal lattice. Nevertheless, structural correlation functions that can be computed by PIGS, such as the pair correlation function (Fig. \ref{gofr}), display the characteristic signs of crystallization as the density is increased. For instance, the pair correlation function displays a main peak that grows stronger as $\rho$ is increased, and secondary peaks also appear.

\section{CONCLUSIONS}
We performed extensive Quantum Monte Carlo simulations of condensed $^4$He at $T$=0, using the Path Integral Ground State method. We utilized a fourth-order approximation for the short imaginary time propagator $G_\circ$, and compared the accuracy and efficiency of this method with other existing techniques, including Diffusion Monte Carlo, Green Function Monte Carlo, Reptation Quantum Monte Carlo as well as Path Integral Ground State with a more accurate form of $G_\circ$, namely the pair product approximation.
  
Our results clearly show that PIGS is a valid alternative to DMC, at least for this particular system; it generally provides more accurate energy results, particularly when the trial wave function used as input to the calculation is only moderately accurate, and qualitatively misses some of the physics (e.g., a translationally invariant wave function, at pressures where the system ought to display crystalline order). We found that, even if the trial wave function does not contain all of the relevant correlations, the projection time needed to extract accurate ground state estimates is relatively small (of the order of 0.25 K$^{-1}$ at the highest density).  In this sense, PIGS seems more robust than DMC. It is worth repeating that PIGS does not suffer from the bias due to the finite population size (affecting both DMC and GFMC), and allows one to compute  expectation values of operators that do not commute with the Hamiltonian more easily than DMC. 

The use of the fourth-order propagator makes it possible to carry out these calculations with a typical number of imaginary time slices $M$=80-160; these simulations are quite feasible on a common desktop workstation, in a reasonable amount of time.
By using the pair product approximation, one may be able to reduce the number of imaginary slices needed by a factor two or so; however,  in our opinion the much greater simplicity and generality of the fourth-order propagator make it a preferable choice. Obviously, ``simplicity"  is a subjective criterion, and opinions may vary. It should also be mentioned that the FOP used here is only one of several possible choices, some of which may be more efficient, possibly closing the gap with the pair product approximation even further.\cite{another_chin}
\label{conclusions}
\section*{Acknowledgments}
This work was supported in part by the Petroleum Research Fund of the American 
Chemical Society under research grant 36658-AC5, and by the Natural Science 
and Engineering Research Council of Canada under research grant G121210893. 
PNR acknowledges The Canada foundation for innovation and the Natural Science 
and Engineering Research Council of Canada for research support.

\bibliography{paper}
\bibliographystyle{jcp}

\newpage

\begin{table}[ht]
\caption{Kinetic ($K$) and total ($e$)  energy per particle (in K) in liquid $^4$He at the density $\rho$=0.0218 \AA$^{-3}$, obtained with different ground state Quantum Monte Carlo methods and with the version of the Aziz potential of Ref. \onlinecite{Aziz-95}. The projection time $\beta$ utilized in both PIGS calculations is 0.4 K$^{-1}$. The third column gives the  number of particles. The GFMC and PIGS-FOP estimates are extrapolated to the $\tau=0$ limit, whereas the PIGS-PPA uses a time step of 0.0125 k$^{-1}$. Statistical errors (in parentheses) are on the last digit(s).}
\begin{ruledtabular}
\begin{tabular}{cccc}
Method & $K$ & $e$  &$N$ \\
GFMC (Ref. \onlinecite {casulleras}) & 13.842 (50) & -7.292(3) & 64 \\
PIGS-PPA (Ref. \onlinecite{Magro})&14.390(60)&-7.318(38) &64\\
PIGS-FOP (This work) &14.241(22) &-7.286(6)  &256 \\
\end{tabular}
\label{tab1}
\end{ruledtabular}
\end{table}

\begin{table}[ht]
\caption{Total energy per particle, $e(\rho)$, in K, computed by PIGS as a function of density, using the trial wave function (\ref{jastrow}) with the pseudo-potential (\ref{pseudo}). 
The potential utilized in this calculation is the Aziz-I.\cite{aziz79} Also shown for comparison are the most recent estimates (for the same potential) obtained with DMC (Ref. \onlinecite{boronat94}) and GFMC (Ref. \onlinecite{kalos}). Statistical errors (in parentheses) are on the last digit(s). Except at the two highest densities, for which $N=288$, PIGS results are obtained with $N=256$ particles.}
\begin{ruledtabular}
\begin{tabular}{llccc}
$\rho$ (\AA$^{-3}$)&PIGS&DMC& GFMC\\ \hline
0.0196 & -7.002(6)& -6.988(13) &-7.034(37)\\
0.0207 &-7.089(7) & & \\
0.0219 &-7.123(3) &-7.121(10) &-7.120(24)\\
0.0240&-6.997(9) &-6.892(13) &-6.894(48)\\
0.0254 & -6.749(9) & -6.696(24) & \\
0.0262 &-6.568(12) & -6.422(20) &-6.564(58) \\
0.0279 &-5.971(15) & & \\
0.0293 &-5.284(17) &-5.010(25) &-5.175(101)\\
\end{tabular}
\label{tab2}
\end{ruledtabular}
\end{table}
\newpage
\centerline{\bf Figure Captions}
\begin{enumerate}
\item{Kinetic  energy per $^4$He atom $K(\tau)$ (in K), versus time step  $\tau$ (in K$^{-1}$). The total projection time is $\beta=0.25$ K$^{-1}$. These calculations were carried out using the Aziz-I potential,\cite{aziz79} at the equilibrium density $\rho_e=0.02186$ \AA$^{-3}$, with  256 particles.  Diamonds show results obtained using the primitive approximations, circles those obtained with the higher-order propagator. 
Dotted and dashed lines are, respectively, a quadratic (primitive approximation) and a quartic (higher-order) fit to the PIGS data. The extrapolated value of the kinetic energy per particle is 14.235$\pm$0.011 K.}
\item{Total   energy per $^4$He atom $e(\tau)$ (in K), versus time step  $\tau$ (in K$^{-1}$). The total projection time is $\beta=0.25$ K$^{-1}$. These calculations were carried out using the Aziz-I potential,\cite{aziz79} at the equilibrium density $\rho_e=0.02186$ \AA$^{-3}$, with  256 particles.  Dashed line is a quartic fit to the PIGS data. The extrapolated value of the total energy per particle is -7.123$\pm$0.003 K.}
\item{Total   energy per $^4$He atom $e(\beta,\tau)$ (in K), versus imaginary projection time  $\beta$ (in K$^{-1}$). The time step used in all calculations is $\tau=3.125\times10^{-3}$ K$^{-1}$. These calculations were carried out using the Aziz-I potential,\cite{aziz79} at the equilibrium density $\rho_e=0.02186$ \AA$^{-3}$, with  256 particles.  Dashed line is a  fit to the PIGS data based on the expression $e(\beta)=a+b\ {\rm exp}(-c\beta)$.
}
\item{(Color online) Comparison of equations of state of condensed $^4$He, computed by PIGS (diamonds) and DMC
(circles, Ref. \onlinecite {boronat94}). Data show the total energy per $^4$He atom $e$ (in K) versus the density 
$\rho$ (in \AA$^{-3}$). Dotted line is a polynomial fit to the diamonds; its only purpose is to guide the eyes.
}
\item{(Color online) Pair correlation function  $g(r)$ for condensed $^4$He, computed by PIGS, at the equilibrium density 
($\rho$=0.02186 \AA$^{-3}$, solid line) and at the melting density ($\rho$=0.0293 \AA$^{-3}$, dotted line). The distance $r$ is in \AA.}
\end{enumerate}
\newpage

\end{document}